\documentclass[a4paper,12pt]{article}
\usepackage{jheppubmod} 
\usepackage{lineno}

\def\0{{\bf 0}}

\def\l{\ell}

\title{\boldmath Entanglement entropy and the boundary action of edge modes}

\author{Jyotirmoy Mukherjee}
\affiliation[a]{ Department of Theoretical Physics\\
Tata Institute for Fundamental Research, Mumbai 400005, India.}
\emailAdd{jyotirmoy.mukherjee\_119@tifr.res.in}

\abstract{We consider an antisymmetric gauge field in the Minkowski space of $d$-dimension and decompose it in terms of the antisymmetric tensor harmonics and fix the gauge. The Gauss law implies that the normal component of the field strength on the spherical entangling surface will label the superselection sectors. From the two-point function of the field strength on the sphere, we evaluate the logarithmic divergent term of the entanglement entropy of edge modes of $p$-form field. We observe that the logarithmic divergent term in entanglement entropy of edge modes coincides with the edge partition function of co-exact $p$-form on the sphere when expressed in terms of the Harish-Chandra characters. We also develop a boundary path integral of the antisymmetric $p$-form gauge field. From the boundary path integral, we show that the edge mode partition function corresponds to the co-exact $(p-1)$-forms on the boundary. This boundary path integral agrees with the direct evaluation of the entanglement entropy of edge modes extracted from the two-point function of the normal component of the field strength on the entangling surface.}

\begin{document}
\maketitle
\flushbottom
\section{Introduction}
Over the last decade entanglement entropy has emerged as a useful quantity to understand the fundamental properties in quantum field theory and quantum gravity. It has been useful to understand the critical phenomenon in condensed matter systems \cite{Vidal:2002rm} as well as in holography \cite{Ryu:2006bv, Ryu:2006ef}. Entanglement entropy is quite easy to define for quantum systems with local degrees of freedom. At a constant time slice, one divides the Hilbert space into two parts $A$ and $\bar{A}$ which means the Hilbert space of the quantum system factorizes into
\begin{align}
    \mathcal{H}&=\mathcal{H}_A\otimes \mathcal{H}_{\bar{A}}.
\end{align}
In the factorized Hilbert space, one can define reduce density matrix by tracing out or integrating out the degrees of freedom in $\mathcal{H}_{\bar{A}}$. Let us denote this by $\rho_A=\rm{Tr}_{\bar{A}}\rho$. The entanglement entropy of the subregion A is defined as
\begin{align}
    S_A&=-\rm{Tr}(\rho_A\log\rho_A).
\end{align}
But, entanglement entropy is not quite easy to define for the theories which admit local gauge symmetry. The major problem in factorization of the Hilbert space comes due to the presence of the non-local gauge invariant operators which do not allow us to factorize the Hilbert space trivially. For gauge theories, the issue is addressed in detail in the literature \cite{Buividovich:2008gq, Donnelly:2011hn, Casini:2013rba, Radicevic:2014kqa, Casini:2014aia, Donnelly:2014gva, Huang:2014pfa, Donnelly:2014fua, Ghosh:2015iwa, Aoki:2015bsa, Donnelly:2015hxa, Soni:2015yga, VanAcoleyen:2015ccp, Casini:2015dsg, Radicevic:2015sza, Soni:2016ogt}. In the operator formalism language, the Hilbert space does not factorize trivially due to the presence of a centre of the algebra of operators. The centre is assigned to the operator which commutes with all other operators in the local algebra. A centre can be chosen in various ways \cite{Casini:2022rlv}, but in the extended Hilbert formalism in lattice gauge theories, a choice of an electric centre naturally arises. The presence of the electric centre comes from the electric Gauss law constraint on the gauge invariant states. However, it is also possible to choose the algebra on lattice gauge theories with the trivial centre and there will not be any entropy associated with it \cite{Casini:2022rlv}.

In this paper, we will evaluate the entropy associated with the electric centre choice of free $p$-forms across a spherical entangling surface. We will also show that the logarithmic coefficient corresponding to the electric centre  of free $p$-forms coincides with the logarithmic divergent piece of the edge partition function of co-exact $p$-forms on a sphere when expressed in terms of the Harish-Chandra characters. A similar property has also been observed in linearized graviton in $d=4$ dimension \cite{David:2021wrw}. We begin with free $p$-forms across a spherical entangling surface. Since gauge invariant states obey the Gauss law and gauge invariant operators which are Wilson loops do not alter the flux across the entangling surface, the Hilbert space $\mathcal{H}_A$ and $\mathcal{H}_{\bar{A}}$ gets decomposed into superselection sectors. Each superselection sector is labelled by the flux configuration across the boundary. In this case, it will be the normal component or the radial component of the field strength and the entanglement entropy will be 
\begin{equation}
    S_A=-\sum_E p_E \log p_E+\sum_E p_E S(\rho^E_A).
\end{equation}
Here $p_E$ is the probability of the flux configuration which labels the superselection sector.  We denote the normal component of the field strength as electric field, i.e, $E\equiv F_{0ri_1\cdots i_{p-1}}$. The first term in this expression is known as classical Shannon entropy and is also referred to as the entanglement
entropy of edge modes. The second term is known as extractable or quantum entanglement entropy because one associates this with the number of Bell pairs extracted from the quantum system \cite{Soni:2016ogt}.

A  direct method of evaluating the entanglement entropy of the edge modes of free $p$-forms involves the construction of probability distribution associated with the superselection sectors. This method was initially developed for free Maxwell field in $d=4$ dimension \cite{Soni:2016ogt} and was later generalized in arbitrary even dimension in \cite{David:2022jfd}. However, in this paper, we adapt a similar approach but our gauge fixing method is different. We first expand the $p$-form gauge potential in terms of the anti-symmetric harmonics, fix the gauge and finally quantize the theory using radial quantization prescription. Using this radial quantization, we find that the radial component of the field strengths is related to the canonical momentum. We show that the entanglement entropy of the edge modes is determined by the two-point function of the radial component of the field strengths on the sphere. Therefore, we explicitly evaluate the two-point function and obtain
\begin{align}
    & \langle 0| F_{0 r\cdots i_{p-1}} (t,  r, \Omega)   F_{0r\cdots i_{p-1}} ( t, r', \Omega') |0\rangle  
 =\nonumber\\
 &\sum_{\ell,\lambda,m} \frac{(\l+p-1)(\l+d-p-2) }{2\pi  (rr')^{d-2}} Q_{\nu_p-\frac{1}{2}}  (\frac{r^2+r'^2}{2rr'})
 H_{i_1\cdots i_{p-1} } (\Omega) H_{i_1\cdots i_{p-1} }(\Omega') ,
\end{align}
where $Q_{\nu_p-\frac{1}{2}}  (\frac{r^2+r'^2}{2rr'})$ is the Legendre function of the second kind and $ H_{i_1\cdots i_{p-1} } (\Omega)$ are the complete antisymmetric tensor harmonics on sphere. The important point is to note that the two-point function is diagonal in the angular momentum basis and the coefficient is the eigenvalues of the Hodge-de Rham Laplacian of co-exact $(p-1)$-form on $S^{d-2}$. It turns out that in the coincident limit, $Q_{\nu_p-\frac{1}{2}}  (\frac{r^2+r'^2}{2rr'})$ is independent of  $\l$ and $p$. Therefore, the leading divergent term is proportional to the eigenvalues of the Hodge-de Rham Laplacian of co-exact $(p-1)$-form on $S^{d-2}$. We explicitly evaluate the entanglement entropy of edge modes of $p$-forms on the sphere and express it as a partition function of the co-exact $(p-1)$-form on the co-dimension-2 sphere. 

We also obtain the boundary action of free $p$-forms on a Lorentzian manifold in $d$-dimension. We follow the path integral prescription in \cite{Blommaert:2018oue, Geiller:2019bti} and obtain the boundary partition function of free $p$-forms across a planar and spherical boundary. We show that the boundary or the edge partition function reduces to the partition function of co-exact $(p-1)$-forms on the boundary. We understand that, in the presence of a boundary, the gauge symmetry of the $p$-form field is broken and the longitudinal parts of the field become dynamical at the boundary. It is important to note that, at a constant time slice, the boundary becomes a co-dimension 2 manifold and therefore the edge partition function obtained from the boundary action precisely coincides with the direction computation of edge mode entropy in the operator formalism.

We study the partition function of free $p$-forms on the sphere. In \cite{David:2021wrw}, it was shown that the free energy of $p$-forms on sphere can be decomposed into the bulk and edge partition function when expressed in terms of its Harish-Chandra characters. We consider the partition function of co-exact $p$-forms on $S^{d}$ and isolate the edge partition function. We find that this edge partition function is nothing but the complete partition function (bulk+edge) of co-exact $(p-1)$-forms on $S^{d-2}$. Therefore, the logarithmic divergent part of the edge partition function on sphere coincides with the logarithmic divergent part of the entanglement entropy of the edge modes of free $p$-forms across a spherical entangling surface. 

The organization of the paper is as follows. In section \eqref{sec2}, we evaluate the entanglement entropy of edge modes across a sphere using operator formalism. We first expand the gauge potential in antisymmetric tensor harmonics, fix the gauge and quantize the theory. Using this mode quantization procedure, we evaluate the two-point function of the normal component of the field strengths which enables us to isolate the entanglement entropy of edge modes. In section \eqref{sec3} and \eqref{sec4}, we obtain the boundary action and edge partition function of free Maxwell field and $p$-forms across a planar and spherical boundary. We show that the edge partition function of $p$-forms is the same as the free energy of $(p-1)$-form on the boundary. Finally, in section \eqref{sec5}, we relate the edge partition function $p$-forms with the entropy of edge modes in the electric centre choice.
\section{Operator formalism}\label{sec2}
In this section, we discuss the method to evaluate the entanglement entropy of edge modes in $U(1)$ gauge theory across a spherical entangling surface. We use the mode quantization procedure of free $p$-forms on the sphere and evaluate the field strengths which label the superselection sectors. In this case, we restrict our attention to even dimensions and obtain the logarithmic divergent part of the entanglement entropy of edge modes in the antisymmetric $p$-form field.

In gauge theories, gauge invariant observables are the Wilson loops and physical states obey the Gauss law constraint. For free $p$-form gauge potential, the Gauss law is given by
\begin{align}
    \nabla_{\mu_1}F^{0\mu_1\cdots\mu_{p}}=0.
\end{align}
This implies that the normal component of the field strengths $F_{0r\cdots\mu_{p}}$ has to match across the spherical entangling surface. Gauge invariant operators or Wilson loops inside the sphere or outside of the sphere will not change the flux across the boundary. As a consequence, the reduced density matrix gets decomposed into superselection sectors, and each sector is labeled by the normal component of the field strengths on the boundary. We denote the normal component of the field strengths $F_{0r\cdots\mu_{p}}$ as $E_{r\cdots \mu_{p}}$ and call it an elctric flux. Therefore the entanglement entropy of edge modes corresponding to the electric centre is given by
\begin{align}
    S_{\rm{edge}}&=-\sum_E p_E \log p_E.
\end{align}
Here $p_E$ is the probability distribution associated with a particular superselection sector.

The $U(1)$ theory is free and therefore $p_E$ is a Gaussian functional of the normal component of the field strengths on the boundary. Here the entangling surface is a sphere and therefore the distribution is given by
\begin{align}
    p_E&=\mathcal{N}\exp\Big[-\frac{1}{2}\int d^{d-2}x d^{d-2}x' F_{0 r\mu_1\cdots\mu_{p-1}}(x)G^{-1}_{rr'}(x,x')F_{0r\mu_1\cdots\mu_{p}}(x')\Big].
\end{align}
Here $x,x'$ are the coordinates on $S^{d-2}$. $G_{rr'}(x,x')$ is the two-point function of the radial component of the field strengths on the sphere.
\begin{align}
    G_{rr;}(x,x')&=\langle0| F_{0r\mu_1\cdots\mu_{p-1}}(x) F_{0r\mu_1\cdots\mu_{p-1}}(x')|0\rangle.
\end{align}
Note that, the two-point function $G_{rr'}$ is bi-tensor but for notational simplicity, we have suppressed the other directions on the sphere.
We now evaluate the classical contribution and obtain
\begin{eqnarray}
S_{{\rm edge} } ( \rho_A)  = - \log {\cal N }+ \int d^{d-2} x d^{d-2} x' 
G_{rr'} ( x, x') G^{-1}_{r'r}( x', x) .
\end{eqnarray}
Note  that the second term involving the integrals is divergent due to the identity given by
\begin{equation}
\int d^{d-2} x d^{d-2} x' 
G_{rr'} ( x, x') G^{-1}_{r'r}( x', x)  = \int d^{d-2} x \delta^{d-2} (0) .
\end{equation}
It is good to regulate the delta function by introducing a cut-off $\epsilon$ which can be thought of as a short distance cut-off along the angular directions on the entangling surface. 
Therefore one can write the following identity
\begin{equation}
\int d^{d-2} x d^{d-2} x' 
G_{rr'} ( x, x') G^{-1}_{r'r}( x', x)  = \frac{ R^{d-2 } {\rm Vol} ( S^{d-2})  }{ \epsilon^{d-2}} .
\end{equation}
Here $R$ is the radius of the  spherical entangling surface.  Note that, this term corresponds to the area divergent term which is not universal because the coefficient would change upon changing the UV cut-off. We wish to obtain the logarithmic divergent part of the entanglement entropy of the edge modes. Therefore we study the normalization term. From the normalizability condition,
\begin{align}
    \int [DE_{r\mu_1\cdots\mu_{p-1}}]p_E=1,
\end{align}
we obtain the following relation
\begin{equation}
\log {\cal N} -  \frac{1}{2} \log\big( {\rm det }\,  G^{-1}_{r r'} \big)  =0.
\end{equation}
Therefore the logarithmic divergent part of the entanglement of the edge modes
is given by 
\begin{equation} \label{unived}
\left. S_{{\rm edge} } ( \rho_A) = \frac{1}{2} \log {\rm det} G_{rr'}\right|_{\log \; {\rm coefficient}}.
\end{equation}
The two-point function of the radial electric field strengths $ F_{0r\mu_1\cdots \mu_{p-1}}$ diverges when field strengths on two different spheres coincide. 
To regulate this divergence we consider the correlator where the radial 
component of the field strengths lie on two spheres of radius $r$ and $ r'= r + \delta$
\begin{equation}\label{splitgrr}
G_{r r' } ( r, r'; x,  x')  = \langle 0|  F_{0 r\mu_1\cdots \mu_{p-1}} ( r, x) F_{0r'\mu_1\cdots\mu_{p-1}} ( r'. x') |0 \rangle.
\end{equation} 
 We obtain the two-point function and replace $r'=r+\delta$. At the end of the computation we take $\delta \rightarrow 0$ and collect the leading order divergent term in order to evaluate the entanglement entropy of edge modes \cite{David:2022czg}.
\subsection{Free Maxwell field }
In this section, we explain the computation of the entanglement entropy of edge modes in the free Maxwell field using a radial quantization prescription developed in \cite{David:2022jfd}. However, we choose a different gauge to quantize the theory. Let us first expand the gauge potential on a spherical harmonic basis.
\begin{align}
    A_0(t,r,\Omega)&=\sum_{\l,\lambda,m}A_{0(\l,\lambda,m)}(t,r)Y_{\l,\lambda,m}(\Omega),\quad  A_r(t,r,\Omega)&=\sum_{\l,\lambda,m}A_{r(\l,\lambda,m)}(t,r)Y_{\l,\lambda,m}(\Omega),\nonumber\\
     A_i(t,r,\Omega)&=\sum_{\l,\lambda,m}A_{i(\l,\lambda,m)}(t,r)Y^i_{\l,\lambda,m}(\Omega).
\end{align}
Here $Y_{\l \lambda m  }$ $Y_{\l \lambda m  }^{{\bf i}}$  are scalar and vector harmonics on $S^{d-2}$ respectively. The detailed properties of the scalar and vector harmonics are given in \cite{Benedetti:2019uej} and   they obey the property 
\begin{equation}
Y_{\l \lambda m }( \Omega)^* = (-1)^m  Y_{\l \lambda - m }( \Omega).
\end{equation}
Due to this relation, one can easily obtain the reality conditions of the gauge potentials \cite{David:2022czg}
\begin{eqnarray} \label{realityd}
A_{0 ( \l , \lambda, m )}^* ( r, t)  = (-1)^m A_{0 ( \l , \lambda, m )} ( r, t) ,  \\ \nonumber
A_{r ( \l , \lambda, m )}^* ( r, t) = (-1)^m A_{r ( \l , \lambda, m )} ( r, t).
\end{eqnarray}
We impose the gauge condition $\nabla_iA^{i}=0$ which means only the transverse components of the gauge fields are  physical on the codimension-2 sphere. Here $A_0$ will serve as a Lagrange multiplier and variation of the action with respect to $A_0$ will give us the constraint that we impose to evaluate the equation of motion. Since we are interested in the two-point functions of the radial components of the field strengths on sphere, we focus our attention on the terms in the Lagrangian which has $A_0$ and $A_r$ gauge potentials.
\begin{align}
S &= \sum_{l \lambda m =-\l}^\l \int dr L_{\l \lambda m } , \nonumber\\
L_{\l \lambda m } &= \frac{r^{d-2}}{2}\left[  \dot A_{r (\l , \lambda, m } \dot  A^*_{r ( \l, \lambda, m ) }
+  \partial_r A_{0 ( \l, \lambda, m ) } \partial_r A_{0 ( \l, \lambda m ) }^* \right. + \frac{\l ( \l + d-3)}{r^2}  A_{r ( \l \lambda m ) }  A_{r ( \l ,\lambda ,m ) } ^* \nonumber\\
& \left. -  \dot A_{r ( \l, \lambda, m )} \partial_r A^*_{0(\l ,\lambda, m) } 
- \dot A^*_{r ( \l, \lambda m ) } \partial_r A_{0 ( \l, \lambda, m ) }
+ \frac{\l ( \l + d-3)}{r^2}  A_{0 ( \l \lambda m ) }  A_{0 ( \l ,\lambda ,m ) } ^* \right].
\end{align}
The canonical conjugate momentum corresponding to $A_r$ mode can be obtained 
\begin{equation} \label{pirdef}
\pi^{r\, *} _{( \l,  \lambda , m ) } = r^{d-2} ( \dot A_{r ( \l, \lambda, m)} - \partial_r A_{0 ( \l, \lambda, m ) } ) . 
\end{equation}
The equation of motion also can be obtained as
\begin{align}\label{eom}
    \dot{\pi}^{r\, *} _{( \l,  \lambda , m ) }+r^{d-4}\l(\l+d-3)A_{r ( \l, \lambda, m)}=0.
\end{align}
The constraint equation can also be obtained as
\begin{equation}
\partial_r \pi^{r\, *}_{( \l, \lambda, m )} +r^{d-4} \l ( \l  + d-3) A_{0 ( \l, \lambda, m ) }=0. 
\end{equation}
We now eliminate $A_0$ and obtain the equation of motion of $A_r$ 
\begin{eqnarray}
\partial_r^2  A_{r ( \l, \lambda, m ) } + \frac{ d-4}{r} \partial_r  A_{r ( \l, \lambda, m ) }
+ \Big(  k^2 - \frac{ \l ( \l + d-3) + d-4}{r^2} \Big)  A_{r ( \l, \lambda, m ) } =0.
\end{eqnarray}
We demand the regularity of the solution at the origin and the solution is determined upto an integration constant $a_{r ( \l, \lambda, m )}$
\begin{equation}\label{solnard}
A_{r ( \l, \lambda, m ) }  =  e^{-i k t} a_{r ( \l, \lambda, m ) } 
  r^{ - \frac{d-5}{2}} J_{\l + \frac{ d-3}{2} } ( |k| r) ,
\end{equation}
From the equation of motion \eqref{eom}, it is easy to determine $\pi^{r\, *} _{( \l,  \lambda , m )}$. We demand the solution to vanish at $r\rightarrow \infty$ which implies
 \begin{eqnarray}\label{solndpi}
 \pi^{r\, *}_{ (\l, \lambda, m ) }  =  - i \frac{\l ( \l + d-3)  }{k } 
 a_{r ( \l, \lambda, m ) } (k) e^{-i k t}     r^{\frac{ d-3}{2}}
 J_{\l + \frac{d -3}{2}} ( |k| r) .
 \end{eqnarray}
  The classical solutions (\ref{solnard}) and (\ref{solndpi}) for each Fourier mode, enable us to write the following mode expansions
\begin{align}\label{modexpd1}
 A_{r ( \l, \lambda, m) } ( r, t ) &= \frac{1}{\sqrt{2} } 
 \int_0^\infty 
  k dk \left( a_{r (\l,  \lambda, m) } (k) e^{ - i k t}  + (-1)^m  a_{r( \l, \lambda, -m )}^\dagger (k)  e^{i k t}  \right) 
  r^{- \frac{d-5}{2} }  J_{\l + \frac{d-3 }{2} } ( k r) ,  \nonumber \\
 \pi^{r\, *}_{( \l,\lambda,  m ) } ( r, t) &=   \frac{ \l ( \l +d-3 ) }{\sqrt{2} } 
 \int_0^\infty  dk \left( - i a_{r (\l, \lambda, m) } (k) e^{ - i k t} +  i (-1)^m a_{r(\l, \lambda,  -m)}^\dagger(k)  e^{i k t}    \right)
 r^{\frac{d-3}{2}}
 J_{\l + \frac{d-3}{2} } ( k r) . 
 \end{align}
 Given the mode expansion of $ \pi^{r\, *}_{( \l,\lambda,  m ) } ( r, t)$, we write down the mode expansion of $\pi^{r\, *}_{( \l,\lambda,  m ) } ( r, t)$
 \begin{eqnarray}\label{modexpd2}
 \pi^{r}_{( \l, m ) } ( r, t) &=&   \frac{ \l ( \l +d-3) }{\sqrt{2} } 
  \int_0^\infty  dk \left(  i a_{r (\l, \lambda, m) }^\dagger (k) e^{  i k t} -  i (-1)^m a_{r(\l,  \lambda, -m)}(k) e^{-i k t} 
    \right) r^{\frac{d-3}{2}} J_{\l + \frac{1}{2} } ( k r) . \nonumber \\
 \end{eqnarray}
The equal time commutation relation of  $A_r, \pi^r$ is given by 
\begin{equation}
[A_{r (\l, \lambda, m ) } (r, t) , \pi_{r ( \l', \lambda', m' ) } ( r, t) ] = i \delta_{\l, \l'} \delta_{\lambda, \lambda'}\delta_{m, m'} \delta( r- r') .
\end{equation}
This implies that creation and annihilation operators should obey the following commutation algebra
\begin{equation}\label{coma}
[a_{r(\l, \lambda, m ) } (k) , a^\dagger_{r(\l', \lambda', m' )} (k')  ] = \frac{1}{\l ( \l + d-3) } \delta_{\l \l'} \delta_{\lambda, \lambda'}\delta_{m, m'}.
\end{equation}
We now evaluate the two-point function of the radial components of the electric field. From the relation \eqref{pirdef}, we identify the relation between electric field operator $F_{0r}$ and canonical conjugate momentum $\pi_r$
\begin{eqnarray}
F_{0  r } &=& \sum_{\l, \lambda, m } \Big ( \dot A_{r ( \l, \lambda, m )} - \partial_r A_{0 ( \l, \lambda, m ) } 
\Big) Y_{(\l\lambda m) } ( \Omega)  , \\ \nonumber
&=& \sum_{\l, \lambda, m }  \frac{ \pi^{r\, *}_{ ( \l, \lambda, m ) } }{ r^{d-2 } }  Y_{(\l\lambda m) } ( \Omega) .
\end{eqnarray}
Using the mode expansion of the canonical conjugate momentum operator $\pi_r$
, 
we  now proceed to evaluate the two-point function of the electric field
\begin{eqnarray}\label{e2}
&& \langle 0| F_{0 r} (t,  r, \Omega)   F_{0r} ( t, r', \Omega') |0\rangle  = 
\\ \nonumber
&& \frac{1}{ 2 ( rr')^{\frac{d-1}{2} } }  \sum_{\l, \l',\lambda,\lambda' m, m'}  \l ( \l +d-3) \l' ( \l' +d-3)  \int_0^\infty dk dk' 
\left[ 
J_{\l + \frac{1}{2} } ( k r)  J_{\l' + \frac{d-3}{2} } ( k' r')    \right. \\ \nonumber
& & \qquad\qquad\qquad\qquad \left. 
 \times  (-1)^{m'} 
 \langle 0| a_{ r (\l, m ) }  (k) a^\dagger_{ r ( \l', m' ) }(k') |0\rangle Y_{\l, m } ( \Omega) Y_{\l', - m'} (\Omega')
 \right].
\end{eqnarray}
Using the commutation relations in (\ref{coma}), we obtain
\begin{eqnarray}\label{e3}
 && \langle 0| F_{0 r} (t,  r, \Omega)   F_{0r} ( t, r', \Omega') |0\rangle  = \\ \nonumber
 && \qquad\qquad
 \frac{1}{ 2 ( rr')^{\frac{d-1}{2} } }\sum_{\l, \lambda,m } \l ( \l +d-3) \int_0^\infty  dk  J_{\l + \frac{d-3}{2} } ( k r)  J_{\l' + \frac{d-3}{2} } ( k' r')  
 Y_{\l, m } ( \Omega) Y_{\l,  m}^* (\Omega'). \nonumber 
\end{eqnarray}
We can now perform the integral using the following identity in  \cite{gradshteyn2007}
\begin{align}\label{besselidentity}
   \int_0^\infty dk  J_{\ell+\frac{1}{2}}(kr)J_{\ell+\frac{1}{2}}(kr')=\frac{1}{\pi\sqrt{rr'}}Q_{\ell}(\frac{r^2+r'^2}{2rr'}),
\end{align}
where $Q_{\nu}(z)$ is the Legendre function of the second kind which can also be expressed in terms of 
the hypergeometric function \cite{gradshteyn2007}
\begin{align}
    Q_{\nu}(z)=\frac{\Gamma \left(\frac{1}{2}\right) \Gamma (\nu +1) z^{-\nu -1} \, _2F_1\left(\frac{\nu +2}{2},\frac{\nu +1}{2};\frac{1}{2} (2 \nu +3);\frac{1}{z^2}\right)}{2^{\nu +1} \Gamma \left(\nu +\frac{3}{2}\right)}.
\end{align}
Therefore the two-point function of the electric field operator is obtained as
\begin{equation}\label{eleccor}
 \langle 0| F_{0 r} (t,  r, \Omega)   F_{0r} ( t, r', \Omega') |0\rangle  
 =\sum_{\ell,m} \frac{\l(\l+d-3) }{2\pi  (rr')^{d-2}} Q_{\ell}  (\frac{r^2+r'^2}{2rr'})
 Y_{\l, m } (\Omega) Y_{\l, m }^*(\Omega') .
\end{equation}
We need to evaluate the two-point function in the coincident limit of the insertions of the electric field operators. Therefore we take
\begin{equation}
r' = r + \delta,  \qquad \delta \rightarrow 0.
\end{equation}
In the coincident limit, the Legendre function of the second kind admits the following expansion \cite{David:2022czg}
\begin{equation}\label{Qexpand}
\lim_{\delta\rightarrow 0} Q_\l \left( \frac{ r^2 + ( r+\delta)^2}{ 2 r ( r+\delta)} \right) 
= \frac{1}{2 } \log( \frac{ r^2}{\delta^2} )  + \log 2 - H_\l + O(\delta) ,
\end{equation}

where $H_\l$ refers to the Harmonic number. 
It is important to note that the leading order divergent term is independent of $\l$.
Substituting this limit in the two-point function (\ref{eleccor}), we obtain 
\begin{equation}
\lim_{\delta\rightarrow 0} \langle 0| F_{0 r} (t,  r, \Omega)   F_{0r} ( t, r', \Omega') |0\rangle   =  \frac{1}{4\pi r^{2(d-2)}}  \log ( \frac{r^2}{\delta^2} )  \sum_{\l \geq 1, m }
\l(\l +d-3) Y_{\l, \lambda,m } ( \Omega) Y_{\l,\lambda, m }^*(\Omega'),
\end{equation}
Here we keep only the leading order divergent term. 
It is clear that the correlator admits a diagonal form in the angular momentum basis and the elements are given by
\begin{equation}\label{fm}
\lim_{\delta\rightarrow 0}\langle \l\lambda|G_{rr'}(r,r+\delta)|\l'\lambda'\rangle = \frac{\l(\l +d-3)}{4\pi r^{2(d-2)}} \log ( \frac{r^2}{\delta^2} )  \delta_{\l, \l'} \delta{\lambda,\lambda'}\delta_{m, m'}, 
\end{equation}
where $G_{rr'}(r,r')$ is the two-point function of the radial components of the electric field on sphere. Note that, the correlator is proportional to the massless scalar Laplacian on $S^{d-2}$. Therefore, the edge mode degrees of freedom across a spherical entangling surface can be identified with the 0-form or the massless scalar on a codimension-2 entangling surface. Now in order to obtain the logarithmic divergent part of the entanglement entropy of edge modes, one requires to obtain the free energy of the massless scalar on $S^{d-2}$.
\begin{eqnarray}
S_{\rm edge} ( \rho_A) = \frac{1}{2} \sum_{l =1}^\infty g_{\l, d} \log ( \l ( \l +  d -3)) ,  \qquad
g_{\l, d} = \frac{  ( 2\l + d - 3) \Gamma( \l + d - 3) }{ \l! \Gamma( d-2) } ,
\end{eqnarray}
where $g_{\l, d}$ are the degeneracies of the Laplacian  of the 0-form or massless scalar on $S^{d-2}$. We perform the sum over the eigenmodes of a scalar field on $S^{d-2}$ and cast the partition function in terms of the Harish-Chandra character
\begin{equation}
-\frac{1}{2} \log \rm{det} (\Delta^{S^{d-2}}_0 ) = \int_0^\infty\frac{dt}{2t} \frac{1 + e^{-t}}{ 1- e^{-t}}  \frac{1 + e^{- ( d-3 ) t} }{ 1- e^{-( d-3) t}} .
\end{equation}
It is important to note that, the partition function is now an integral transform of Harish-Chandra characters of 0-form or massless scalar on $S^{d-2}$. The Harish-Chandra character of massless scalar on $S^d$ is given by \cite{Anninos:2020hfj, David:2021wrw}
\begin{equation}
\chi_{ d, 0}^{\rm dS}(t) = \frac{1 + e^{- (  d -1 ) t} }{ 1- e^{-(  d - 1) t}} .
\end{equation}
\subsection{2-form}
In the previous section, we derive the entanglement entropy of edge modes of $1$-form across a spherical entangling surface and identify the edge modes with the massless scalar field on the entangling surface.
In this section, we generalize the computation in $p$-forms and understand the edge mode degrees of freedom. We evaluate the entanglement entropy of these edge modes in Minkowski space of $d$dimesion across a spherical entangling surface.
The metric is given by
\begin{align}
    ds^2=-dt^2+dr^2+r^2d\Omega_{d-2}^2.
\end{align}
$d\Omega_{d-2}^2$ represents the metric on $S^{d-2}$. Using the spherical symmetry of the problem we first quantize $p$-forms using the radial quantization procedure. We first show this explicitly in $2$-forms and later we will generalize it to arbitrary $p$-forms. The action of the $ 2$ form is given by
\begin{align}
    \mathcal{L}&=-\frac{1}{12}\int r^{d-2}drd\Omega_{d-2}F_{\mu\nu\rho}F^{\mu\nu\rho},\nonumber\\
    &=-\frac{1}{4}\int r^{d-2}drd\Omega_{d-2}\left(F_{0ri}F^{0ri}+F_{rij}F^{rij}+F_{0ij}F^{0ij}+\frac{1}{3}F_{ijk}F^{ijk}\right), \quad\quad i\in S^{d-2}\nonumber\\
    &=\mathcal{L}_1+\mathcal{L}_2.
\end{align}
Here $\mathcal{L}_1$ and $\mathcal{L}_2$ are given by
\begin{align}
    \begin{split}
        \mathcal{L}_1&=-\frac{1}{4}\int r^{d-2}drd\Omega_{d-2}\left(F_{0ri}F^{0ri}+F_{rij}F^{rij}+F_{0ij}F^{0ij}\right),\\
        \mathcal{L}_2&=-\frac{1}{12}\int r^{d-2}drd\Omega_{d-2} F_{ijk}F^{ijk}.
    \end{split}
\end{align}
Since superselection sectors are labeled by the $F_{0ri}$ field strength, we focus our attention to $\mathcal{L}_1$ to obtain two-point function of $F_{0ri}.$  We first expand the vector potential $A_{\mu\nu}$ in terms of spherical harmonics.
\begin{align}
 A_{0r}(t,r,\Omega)&=\sum_{\l,\lambda,m}A_{0r(\l,\lambda,m)}(t,r)H^{0r}_{\l,\lambda,m}(\Omega),\quad  A_{0i}(t,r,\Omega)&=\sum_{\l,\lambda,m}A_{0i(\l,\lambda,m)}(t,r)H^{0i}_{\l,\lambda,m}(\Omega)\nonumber\\
 A_{ri}(t,r,\Omega)&=\sum_{\l,\lambda,m}A_{ri(\l,\lambda,m)}(t,r)H^{ri}_{\l,\lambda,m}(\Omega),\quad  A_{ij}(t,r,\Omega)&=\sum_{\l,\lambda,m}A_{ij(\l,\lambda,m)}(t,r)H^{ij}_{\l,\lambda,m}(\Omega).
\end{align}
Here $\{i,j\}\in S^{d-2}$ and $ H^{\mu,\nu}_{\l \lambda m, }$ are the antisymmetric tensor harmonics. Since the entangling surface is a sphere of fixed radius $R$, the tensor harmonics $H^{ri}_{(\l,\lambda,m)}$ is actually a vector harmonics and $H^{0r}_{(\l,\lambda,m)}$ is a scalar harmonics.

We also impose the following gauge condition
\begin{align}
   \quad\quad \nabla_iA^{ij}=0.
\end{align}
This means only transverse modes on the sphere are physical degrees of freedom. We will also have $A_{0r}$ and $A_{0i}$ gauge potentials which act as Lagrange multipliers. To evaluate the two-point function of $F_{0ri}$, we will require $\mathcal{L}_1$ and only those terms involving kinetic terms of $A_{ri}$ and Lagrange multipliers.

Let us now focus on the relevant part of the Lagrangian which consists of the kinetic and potential terms of $A_{ri}$ as well as the Lagrange multipliers.
\begin{align}
   \mathcal{L}'_{\l,\lambda,m}&=\frac{r^{d-2}}{4}\Big[\dot{A}_{ri,(\l,\lambda,m)}\dot{A}_{r,(\l,\lambda,m)}^{*i}+\nabla_r A_{i0,(\l,\lambda,m)}\nabla_r A^{*i}_{~~0,(\l,\lambda,m)}+\dot{A}_{r,(\l,\lambda,m)}^{*i}\nabla_r A_{i0,(\l,\lambda,m)}\nonumber\\
   &+\dot{A}_{r,(\l,\lambda,m)}^{*i}\nabla_i A_{0r,(\l,\lambda,m)}+\frac{(\l+1)(\l+d-4)}{r^2} A^*_{0r,(\l,\lambda,m)} A_{0r,(\l,\lambda,m)}+\dot{A}_{r,(\l,\lambda,m)}^{~i}\nabla_r A^*_{i0,(\l,\lambda,m)}\nonumber\\
   &+\dot{A}_{r,(\l,\lambda,m)}^{~i}\nabla_i A^*_{0r,(\l,\lambda,m)}-\frac{(\l+1)(\l+d-4)}{r^2} A_{r,(\l,\lambda,m)}^{*i} A_{ri,(\l,\lambda,m)}\Big].
\end{align}
From this part of the Lagrangian, it is easy to evaluate canonical conjugate momentum.
\begin{align}\label{pirdef2f}
 \pi_{ri,(\l,\lambda,m)}=  \frac{\partial L_{1a}}{\partial \dot A_{ r ( \l ,\lambda, m ) }^{*i} }&=\frac{r^{d-2}}{4}\left(\dot A_{ri (\l , \lambda, m }+\nabla_r A_{i0(\l ,\lambda, m) }+\nabla_i A_{0r(\l ,\lambda, m)}\right).
\end{align}
It is simple to obtain the equation of motion of the gauge potential $A_{ri}$. This is given by
\begin{align}\label{eom21}
    \dot{\pi}_{ri} +r^{d-4}(\l+1)(\l+d-4)A_{ri}=0.
\end{align}

The gauge potentials $A^*_{0r(\l,\lambda,m)}$ and $A^*_{0i(\l,\lambda,m)}$ are the Lagrangian multiplier and we obtain constraint equations by varying $\mathcal{L}'$ with respect to the gauge potentials $A_{0r(\l,\lambda,m)}$ and $A_{0i(\l,\lambda,m)}$. We first vary the Lagrangian with respect to $A^*_{i0(\l,\lambda,m)}$ and obtain
\begin{align}\label{cons2}
    -\nabla_r\pi_{ri}+  r^{d-4}(\l+1)(\l+d-4)A_{i0}=0.
\end{align}
Using constraint equation \eqref{cons2}, the equation of motion becomes
\begin{align}
    \left(\ddot A_{ri }-\nabla_r\left(\frac{1}{r^{d-4}}\nabla_r(r^{d-4}A_{ri})\right)\right)+\frac{(\l+1)(\l+d-4)}{r^2}A_{ri}=0
\end{align}
We now expand $A_{ri}$ in Fourier modes to solve the equation of motion and obtain
\begin{align}
    \partial_r^2A_{ri}-\frac{2}{r}\partial_rA_{ri}-\frac{1}{r^2}\left((\l+1)(\l+d-4)+\frac{d(d-6)}{4}\right)A_{ri}+k^2A_{ri}=0.
\end{align}
We demand the regularity of the solution at the origin and we obtain
\begin{align}\label{2forma}
    A_{ri(\l,\lambda,m)}&=e^{-ikt}a_{ri(\l,\lambda,m)}r^{\frac{3}{2}}J_{\nu}(|k|r),\quad\quad\nu=\sqrt{(\l+\frac{d-3}{2})^2-(d-4)}.
\end{align}
Using the equation of motion in \eqref{eom21}, we can obtain the solution of the canonical conjugate momentum
\begin{align}\label{2formp}
    \pi^*_{ri(\l,\lambda,m)}&=-\frac{i}{k}(\l+1)(\l+d-4)e^{ikt}r^{d-\frac{5}{2}}a_{ri(\l,\lambda,m)}r^{\frac{3}{2}}J_{\nu}(|k|r),\quad\nu=\sqrt{(\l+\frac{d-3}{2})^2-(d-4)}.
\end{align}
 The classical solutions (\ref{2forma}) and (\ref{2formp}) for each Fourier mode, enable us to write the following mode expansions
\begin{align}\label{modexpd2}
 A_{ri ( \l, \lambda, m) } ( r, t ) &= \frac{1}{\sqrt{2} } 
 \int_0^\infty 
  k dk \left( a_{ri (\l,  \lambda, m) } (k) e^{ - i k t}  + (-1)^m  a_{ri( \l, \lambda, -m )}^\dagger (k)  e^{i k t}  \right) 
  r^{\frac{3}{2} }  J_{\nu } ( k r) ,  \nonumber\\
 \pi^*_{ri( \l,\lambda,  m ) } ( r, t) &=   \frac{ (\l+1) ( \l +d-4 ) }{\sqrt{2} } 
 \int_0^\infty  dk \Big( - i a_{ri (\l, \lambda, m) } (k) e^{ - i k t}\nonumber\\
 &\quad\quad\quad\quad\quad\quad\quad\quad\quad\quad\quad\quad\quad+  i (-1)^m a_{ri(\l, \lambda,  -m)}^\dagger(k)  e^{i k t}    \Big)
 r^{d-\frac{5}{2}}
 J_{\nu } ( k r).
 \end{align}
 Given the mode expansion of $ \pi^*_{ri( \l,\lambda,  m ) } ( r, t)$, we write down the mode expansion of $\pi_{ri( \l,\lambda,  m ) } ( r, t)$
 \begin{align}\label{modexpd2}
 \pi_{ri( \l, \lambda,m ) } ( r, t) &=  \frac{ (\l+1) ( \l +d-4 ) }{\sqrt{2} } 
 \int_0^\infty  dk \Big( i a_{ri (\l, \lambda, m) } (k) e^{ - i k t}\nonumber\\
 &\quad\quad\quad\quad\quad\quad\quad\quad\quad\quad\quad\quad\quad- i (-1)^m a_{ri(\l, \lambda,  -m)}^\dagger(k)  e^{i k t}    \Big)
 r^{d-\frac{5}{2}}
 J_{\nu } ( k r) . 
 \end{align}
 The equal time commutation relation of the fields $A_{ri}, \pi_{ri}$ is given by 
\begin{equation}
[A_{ri (\l, \lambda, m ) } (r, t) , \pi_{ri ( \l', \lambda', m' ) } ( r, t) ] = i \delta_{\l, \l'} \delta_{\lambda, \lambda'}\delta_{m, m'} \delta( r- r') .
\end{equation}
This implies that creation and annihilation operators should obey the following commutation algebra
\begin{equation}\label{coma}
[a_{ri(\l, \lambda, m ) } (k) , a^\dagger_{ri(\l', \lambda', m' )} (k')  ] = \frac{1}{(\l+1) ( \l + d-4) } \delta_{\l \l'} \delta_{\lambda, \lambda'}\delta_{m, m'}.
\end{equation}
We now need to evaluate the two-point function of the radial components of the field strengths. From the definition \eqref{pirdef2f}, we identify the relation between field strength operator $F_{0ri}$ and canonical conjugate momentum $\pi_{ri}$
\begin{eqnarray}
F_{0  r i} 
&=& \sum_{\l, \lambda, m }  \frac{ \pi^*_{ri ( \l, \lambda, m ) } }{ r^{d-2 } }  H^{ri}_{\l\lambda m } ( \Omega) .
\end{eqnarray}
Using the mode expansion of the momentum operator $\pi_{ri}$ and following steps given in \eqref{e2} to \eqref{e3}, we obtain the two-point function of the field strengths
\begin{equation}\label{eleccor2f}
 \langle 0| F_{0 ri} (t,  r, \Omega)   F_{0ri} ( t, r', \Omega') |0\rangle  
 =\sum_{\ell,\lambda,m} \frac{(\l+1)(\l+d-4) }{2\pi  (rr')^{d-2}} Q_{\nu-\frac{1}{2}}  (\frac{r^2+r'^2}{2rr'})
 H_{ri(\l, m) } (\Omega) H^*_{ri(\l, m) }(\Omega') .
\end{equation}
Here $H_{ri(\l, m) } (\Omega)$ is the vector harmonics on $S^{d-2}$ satisfies the following eigenvalue equation
\begin{align}
    -\Delta_{\rm{HdR}}H_{ri(\l, m) } (\Omega)=(\l+1)(\l+d-4)H_{ri(\l, m) } (\Omega).
\end{align}
It is explicit that the two-point function is diagonal in the angular momentum basis and the leading order divergent term is independent of $\l$. The important point is to note here that, the leading order divergent term in the coincident limit of the two-point function is proportional to the eigenvalue of the Hodge-de Rham Laplacian of 1-form. Therefore, the edge modes of the 2-form on sphere can be identified with the co-exact 1-form on the entangling surface. The logarithmic divergent piece of the entanglement entropy of edge modes of 2-form can be extracted by evaluating
\begin{align}\label{2edge}
    S_{\rm{edge}}(\rho_A)&=-\frac{1}{2}\log\det\Delta^{S^{d-2}}_{(1)\rm{HdR}},
\end{align}
where $\Delta^{S^{d-2}}_{(1)\rm{HdR}}$ is the Hodge-de Rham Laplacian of co-exact 1-form on $S^{d-2}$. The logarithmic divergent part of the entanglement entropy of edge mods in 2-form given in  \eqref{2edge} can be evaluated in different ways. We proceed with the heat-kernel method and express it in terms of the Harish-Chandra characters of co-exact 1-form \cite{David:2021wrw}.
\begin{align}
S_{\rm edge} ( \rho_A) &= \frac{1}{2} \sum_{l =1}^\infty g^{(1)}_{\l} \log (( \l+1) ( \l +  d -4)) ,  \quad
g^{(1)}_{\l}& = \frac{(2\l+d-3)\Gamma[\l+d-2]}{\Gamma[d-3](\l-1)!(\l+1)(\l+d-4)} ,
\end{align}
where $g^{(1)}_{\l, d}$ are the degeneracies of the Laplacian  of the co-exact 1-form on $S^{d-2}$. We perform the sum over the eigenmodes of the 1-form field on $S^{d-2}$ and cast the partition function in terms of the Harish-Chandra character \cite{David:2021wrw}
\begin{equation}\label{HdR1}
-\frac{1}{2} \log \rm{det} (\Delta^{S^{d-2}}_{(1)\rm{HdR}} ) = \int_0^\infty\frac{dt}{2t} \frac{1 + e^{-t}}{ 1- e^{-t}}   \left( (d-3)\frac{ e^{-( d-4) t } + e^{-t} }{ ( 1- e^{-t} )^{d-3} 
  }-\frac{1+e^{-(d-5)t}}{(1-e^{-t})^{d-5}} \right). 
\end{equation}
The partition function is expressed in terms of `naive' characters in the sense of \cite{Anninos:2020hfj} and one has to flip it in order to obtain the character corresponding to UIR of $SO(1,d-2)$. However, the log divergent term remains invariant under this flipping procedure and can be extracted from the expansion of the integrand about $t=0$ and collecting the coefficient of $1/t$ term.
\subsection{Generalization to $p$-forms}\label{sec2p}
The explicit computation of $2$-form mode quantization will help us to generalize it to free $p$-form gauge potential on the sphere. The action of free $p$-form gauge potential is given by
\begin{align}
    \mathcal{L}&=-\frac{1}{2(p+1)!}\int r^{d-2}dr d\Omega_{d-2}F_{\mu_1\cdots\mu_{p+1}}F^{\mu_1\cdots\mu_{p+1}}.
\end{align}
Since we require to evaluate the two-point function radial component of the field strengths, we consider only those terms in the Lagrangian which has the kinetic and potential terms of the radial components of the gauge potential. Here $A_{0r i_1\cdots i_{p-2}}$ and $A_{0 i_1\cdots i_{p-1}}$ will serve as Lagrange multipliers. The relevant terms in the Lagrangian are given by
\begin{align}
 S'&=-\frac{1}{2(p+1)!}\int r^{d-2}dr d\Omega\Big[F_{0ri_1\cdots i_{p-1}} F^{0ri_1\cdots i_{p-1}}+F_{0i_1\cdots i_{p}}F^{0 i_1\cdots i_{p}}+F_{ri_1\cdots i_{p}}F^{ri_1\cdots i_{p}}].
\end{align}
 From the action, it is easy to find out the canonical conjugate momentum of the gauge potential $A_{r i_1\cdots i_{p-1}}$
 \begin{align}
      \pi_{ri_1\cdots i_{p-1}}&=\frac{\partial \mathcal{L}}{\partial \dot{A}_r^{~i_1\cdots i_{p-1}}}\nonumber\\
      &=\frac{r^{d-2}}{2}\Big[\dot{A}_{r i_1\cdots i_{p-1}}+\rm{cyclic}].
 \end{align}
 From the definition of the field strengths, we identify
 \begin{align}\label{defc}
     F_{0r i_1\cdots i_{p}}&=\sum_{\l,\lambda,m}\frac{\pi_{ri_1\cdots i_{p-1}}}{r^{d-2}}H^{ri_1\cdots i_{p-1}}_{\l,\lambda,m}(\Omega).
 \end{align}
 Here $H^{ri_1\cdots i_{p-1}}$ are antisymmetric spherical harmonics. The expressions of these harmonics and properties are given in \cite{Camporesi:1994ga}.
 
It is now simple to obtain the equation of motion of the gauge potential $A_{ri_1\cdots i_{p-1}}$. This is given by
\begin{align}\label{eom2}
    \dot{\pi}_{ri_1\cdots i_{p-1}} +r^{d-4}(\l+p-1)(\l+d-p-2)A_{ri_1\cdots i_{p-1}}=0.
\end{align}
The constraint equation can be obtained by varying the action with respect to $A_{0 i_1\cdots i_{p-1}}.$
\begin{align}
     -\nabla_r\pi_{ri_1\cdots i_{p-1}}+  r^{d-4}(\l+p-1)(\l+d-p-2)A_{i_1\cdots i_{p-1}0}=0.
\end{align}
Using this constraint, we can eliminate the Lagrange multipliers from the equation of motion which finally becomes
\begin{align}
    \left(\ddot A_{ri_1\cdots i_{p-1} }-\nabla_r\left(\frac{1}{r^{d-4}}\nabla_r(r^{d-4}A_{ri_1\cdots i_{p-1}})\right)\right)+\frac{(\l+p-1)(\l+d-p-2)}{r^2}A_{ri_1\cdots i_{p-1}}=0.
\end{align}
Expanding the gauge field $A_{ri_1\cdots i_{p-1}}$ in Fourier modes, we solve the equation of motion
\begin{align}
   & \partial_r^2 A_{ri_1\cdots i_{p-1}}+\frac{d-4-(d-2)(p-1)}{r}\partial_r A_{ri_1\cdots i_{p-1}}\nonumber\\
    &+\left(k^2-\frac{(l+p-1) (d+l-p-2)}{r^2}+\frac{(d (p-1)-2 p+4) ((d-2) p-3 d+10)}{4 r^2}\right)A_{ri_1\cdots i_{p-1}}=0.
\end{align}
The solution to this equation is given by
\begin{align}
    A_{ri_1\cdots i_{p-1}}&=e^{-ikt}a_{ri_1\cdots i_{p-1}}r^{\frac{3+(p-2)(d-2)}{2}}J_{\nu_p}(|k|r),\quad\nu_p=\sqrt{\left(\l+\frac{d-3}{2}\right)^2-(p-1)(d-p-2)}.
\end{align}
To obtain the classical solution of the gauge potential $ A_{ri_1\cdots i_{p-1}}$, we just demand the regularity at the origin. Therefore, we have an integral constant in the expression. From the equation of motion, it is easy to obtain the canonical conjugate momentum 
\begin{align}
    \pi^*_{ri_1\cdots i_{p-1}}&=-\frac{i(\l+p-1)(\l+d-p-2)}{k}e^{-ikt}a_{ri_1\cdots i_{p-1}}r^{\frac{p(d-2)-1}{2}}J_{\nu_p}(|k|r).
\end{align}
For notational simplicity, we denote $A_{ri_1\cdots i_{p-1}}\equiv A_{rp}$ and $\pi_{ri_1\cdots i_{p-1}}\equiv \pi_{rp}$. Given the equation of motion of the gauge potential $A_{rp}$ and its conjugate momentum $\pi_{rp}$, we write the mode expansion
\begin{align}\label{modexpd2}
 A_{rp ( \l, \lambda, m) } ( r, t ) &= \frac{1}{\sqrt{2} } 
 \int_0^\infty 
  k dk \left( a_{rp (\l,  \lambda, m) } (k) e^{ - i k t}  + (-1)^m  a_{rp( \l, \lambda, -m )}^\dagger (k)  e^{i k t}  \right) 
  r^{\frac{3+(p-2)(d-2)}{2} }  J_{\nu_p } ( k r) ,  \nonumber\\
 \pi^*_{rp( \l,\lambda,  m ) } ( r, t) &=   \frac{ (\l+p-1) ( \l +d-p-2 ) }{\sqrt{2} } 
 \int_0^\infty  dk \Big( - i a_{rp (\l, \lambda, m) } (k) e^{ - i k t}\nonumber\\
 &\quad\quad\quad\quad\quad\quad\quad\quad\quad\quad\quad\quad\quad\quad+  i (-1)^m a_{rp(\l, \lambda,  -m)}^\dagger(k)  e^{i k t}    \Big)
 r^{\frac{p(d-2)-1}{2}}
 J_{\nu_p } ( k r).
 \end{align}

  Given the mode expansion of $ \pi^*_{rp( \l,\lambda,  m ) } ( r, t)$, we write down the mode expansion of $\pi_{rp( \l,\lambda,  m ) } ( r, t)$
 \begin{align}\label{modexpd2}
 \pi_{rp( \l, \lambda,m ) } ( r, t) &=  \frac{ (\l+p-1) ( \l +d-p-2) }{\sqrt{2} } 
  \int_0^\infty  dk \Big(  i a_{rp (\l, \lambda, m) }^\dagger (k) e^{  i k t} \nonumber\\
  &-  i (-1)^m a_{rp(\l,  \lambda, -m)}(k) e^{-i k t} 
    \Big) r^{\frac{p(d-2)-1}{2}} J_{\nu_p } ( k r) . 
 \end{align}
 The equal time commutation relation of the fields $A_{rp}, \pi_{rp}$ is given by 
\begin{equation}
[A_{rp (\l, \lambda, m ) } (r, t) , \pi_{rp ( \l', \lambda', m' ) } ( r, t) ] = i \delta_{\l, \l'} \delta_{\lambda, \lambda'}\delta_{m, m'} \delta( r- r') .
\end{equation}
This implies that creation and annihilation operators should obey the following commutation algebra
\begin{equation}\label{coma}
[a_{rp(\l, \lambda, m ) } (k) , a^\dagger_{rp(\l', \lambda', m' )} (k')  ] = \frac{1}{(\l+p-1) ( \l + d-p-2) } \delta_{\l \l'} \delta_{\lambda, \lambda'}\delta_{m, m'}.
\end{equation}
We now need to evaluate the two-point function of the radial components of the field strengths. From the definition \eqref{defc}, we identify the relation between field strength operator $F_{0ri_1\cdots i_{p-1}}$ and canonical conjugate momentum $\pi_{ri_1\cdots i_{p-1}}$

Using the mode expansion of the momentum operator $\pi_{ri_1\cdots i_{p-1}}$ and following steps in \eqref{e2} and \eqref{e3}, we obtain the two-point function
\begin{align}\label{eleccor2f}
& \langle 0| F_{0 ri_1\cdots i_{p-1}} (t,  r, \Omega)   F_{0ri_1\cdots i_{p-1}} ( t, r', \Omega') |0\rangle  
 =\nonumber\\
 &\sum_{\ell,\lambda,m} \frac{(\l+p-1)(\l+d-p-2) }{2\pi  (rr')^{d-2}} Q_{\nu_p-\frac{1}{2}}  (\frac{r^2+r'^2}{2rr'})
 H_{ri_1\cdots i_{p-1} } (\Omega) H_{ri_1\cdots i_{p-1} }(\Omega') .
\end{align} 
Similar to 1-form and 2-form, the leading order divergent term in the coincident limit of the two-point function becomes proportional to the Laplacian of Hodge-de Rham Laplacian of co-exact $(p-1)$-form on $S^{d-2}$. Therefore, the entanglement entropy of edge modes corresponding to the electric center of $p$-forms across a spherical entangling surface can be obtained
\begin{align}\label{pform}
    S_{\rm{edge}}&=-\frac{1}{2}\log\det \Delta^{S^{d-2}}_{(p-1)\rm{HdR}.}
\end{align}
It is important to mention that our radial quantization method to evaluate the entanglement entropy of edge modes across a spherical entangling surface agrees with the computation of entanglement entropy of edge modes of free $p$-forms across a planar boundary \cite{Moitra:2018lxn}.
We  now evaluate the logarithmic divergent part of the edge mode entropy by evaluating \eqref{pform}. We cast this in terms of the characters of the co-exact $(p-1)$-form \cite{David:2021wrw}.
\begin{align}
S_{\rm edge} ( \rho_A) &= \frac{1}{2} \sum_{l =1}^\infty g^{(p-1)}_{\l} \log (( \l+p-1) ( \l +  d -p-2)),\nonumber\\
g_{\l}^{(p-1)} &=\frac{(2\l+d-3)\Gamma[\l+d-2]}{\Gamma[p]\Gamma[d-p-4]\Gamma[\l](\l+p-1)(\l+d-p-4)}.
\end{align}
Here $g_{\l}^{(p-1)}$ is the degeneracy of co-exact $(p-1)$-form on $S^{d-2}$.
This sum can be performed and the partition function of co-exact $(p-1)$-form can be expressed as integral over characters \cite{David:2021wrw}.
\begin{align}\label{ppart}
    S_{\rm{edge}}&=\int_{0}^{\infty}\frac{dt}{2t}\frac{ 1+ e^{-t}}{ 1- e^{-t} }
  \left(\sum_{i=0}^{p-1}(-1)^i \chi^{dS}_{(d-3-2i,p-1-i)}(t)\right), 
\end{align}
where $\chi^{dS}_{(d-3-2i,p-1-i)}(t)$ is the naive character in the sense of \cite{Anninos:2020hfj} of co-exact $(p-1)$-form on $S^{d-2}$. The expression of the character of co-exact $p$-form on $S^{d+1}$ was obtained in \cite{David:2021wrw}
\begin{align}
    \chi^{dS}_{(d,p)}(t) &=\binom{d}{p}\frac{e^{-t(d-p)}+e^{-tp}}{(1-e^{-t})^{d}}.
\end{align}
The logarithmic divergent part of the integral \eqref{ppart} can be obtained by expanding the integrand around $t=0$ and collecting the coefficient of $1/t$ term of the integrand.
\section{Edge mode action in free Maxwell theory}\label{sec3}
In this section, we derive the action for the edge modes in the free Maxwell field in a Minkowski space of $d$-dimension. In \cite{Blommaert:2018oue, Geiller:2019bti} it has been shown that the boundary action of the free Maxwell field on the manifold $\mathcal{R}\cup \bar{\mathcal{R}}$ involves the boundary gauge field and the source at the boundary. From this boundary action, we will show the edge modes of the free Maxwell field correspond to the massless scalar degrees of freedom at the boundary.

Let us consider free Maxwell theory on the manifold $\mathcal{R}\cup \bar{\mathcal{R}}$. The action is given by
\begin{equation}
S=-\frac{1}{4}\int _{\mathcal{R}\cup \bar{\mathcal{R}}}d^d x\sqrt{-g}\,F^{\mu\nu}F_{\mu\nu}.
\end{equation}
Let us define two separate gauge fields $A$ and $\bar{A}$, on the manifolds  $\mathcal{R}$ and $\bar{\mathcal{R}}$ respectively. The full path integral can be split into two parts and glued together across the boundary $\partial\mathcal{R}$ by imposing a delta function constraint
\begin{equation}
\int \left[\mathcal{D} A_\mu\right]\left[\mathcal{D} \bar{A}_\mu \right] \prod_{x \in \partial \mathcal{R}} \delta\left(A_{\mu}-\bar{A}_{\mu}\right) e^{i S_A} e^{i S_{\bar{A}}}.
\end{equation}
This delta-function constraint can be expressed as a path integral over boundary currents $j^{\mu}$ in the following way
\begin{equation}
\prod_{x \in \partial \mathcal{R}} \delta\left(A_\mu-\bar{A}_\mu\right) = \int \left[\mathcal{D}j^\mu\right] e^{i\int_{\partial \mathcal{R}}d^{d-1}x\, j^\mu ( A_\mu-\bar{A}_\mu)}.
\end{equation}
Therefore, total action is given by
\begin{align}
\nonumber S= S_A + S_{\bar{A}} =&-\frac{1}{4}\int _{\mathcal{R}}d^d x\sqrt{-g}\,F^{\mu\nu}F_{\mu\nu}\\&+\int_{\partial \mathcal{R}}d^{d-1}x\, j ^\mu ( A_\mu-\bar{A}_\mu)-\frac{1}{4}\int _{\bar{\mathcal{R}}}d^d x\sqrt{-g}\,\bar{F}^{\mu\nu}\bar{F}_{\mu\nu}.
\end{align}
We now have to perform path integral over $A_\mu,\bar{A}_\mu$ and the boundary current $j^{\mu}$. Note that, the boundary current $j^{\mu}$ has been introduced to implement the delta-function constraint along the boundary and hence it serves as a Lagrange multiplier. Therefore, there is no reason for it to be conserved in the first place. Also, note that there is no free charge in the bulk.

Let us now focus on the action of the gauge field $A_{\mu}$
\begin{align}\label{actA}
    S_A&=-\frac{1}{4}\int _{\mathcal{R}}d^d x\sqrt{-g}\,F^{\mu\nu}F_{\mu\nu}+\int_{\partial \mathcal{R}}d^{d-1}x\, j ^\mu A_\mu.
\end{align}
We decompose the gauge field into transverse $A^{\perp}_{\mu}$ and longitudinal modes $\phi$. 
\begin{align}
    A_{\mu}&=A^{\perp}_{\mu}+\partial_{\mu}\phi.
\end{align}
Under the gauge transformation $A_{\mu}\rightarrow A_{\mu}+\partial_{\mu}\epsilon$, the transverse part does not change but the longitudinal part gets a constant shift.
\begin{align}
    A^{\perp}_{\mu}\rightarrow  A^{\perp}_{\mu},\quad\quad \phi\rightarrow \phi+\epsilon.
\end{align}
Note that, the first part of the action \eqref{actA} is manifestly gauge-invariant but the second part is not. Therefore, the action \eqref{actA} becomes
\begin{align}\label{mwactiontotal}
    S_A&=-\frac{1}{4}\int _{\mathcal{R}}d^d x\sqrt{-g}\,F^{\mu\nu}F_{\mu\nu}+\int_{\partial \mathcal{R}}d^{d-1}x\, j ^\mu ( A^{\perp}_\mu+\partial_{\mu}\phi).
\end{align}
It is clear that in the presence of a boundary, the gauge invariance is broken and the longitudinal mode $\phi$ along the boundary becomes a physical field. To evaluate the partition function, we have to path integrate over the transverse gauge field $A^{\perp}_{\mu}$, longitudinal mode $\phi$, and the boundary current $j^{\mu}$. Path integral over transverse gauge field $A^{\perp}_{\mu}$ will give us the bulk partition function which can be identified with the  spin-1 bulk degrees of freedom.
\begin{align}
    \mathcal{Z}[j,\phi]&=\int [D A^{\perp}_{\mu}][D\phi]e^{i S_A}=\mathcal{Z}_{\rm{bulk}}\times \mathcal{Z}_{\rm{bdy}}[j,\phi], \quad Z_{\rm{bulk}}=\frac{1}{\sqrt{\rm{det}\mathcal{O}^{(1)}}}.
\end{align}
To isolate the boundary partition function, we strip off the bulk part and path integrate over the boundary action where boundary gauge field $A^{\perp}_{\mu}$ is evaluated on-shell. Note that, for the boundary action, we can not set the boundary fields to zero rather we use the following boundary condition along the tangential direction of the boundary \cite{Blommaert:2018rsf, Geiller:2019bti}
\begin{equation}\label{bdyc}
\left.\left(\sqrt{-g} n_\mu F^{\mu\alpha}\right)\right|_\text{bdy}= j^\alpha, \quad\quad n_{\alpha}A^{\perp\alpha}|_{\rm{bdy}}=0
\end{equation}
Here $x_{\alpha}$ represents only boundary coordinates and there is no component of the current normal to the boundary, i.e., $j^{n}=0$. Due to the boundary conditions \eqref{bdyc}, we find that there is an isomorphism between the boundary current $j_{\alpha}$ and $A^{\perp}_{\alpha}$. Therefore, for the transverse gauge field $A^{\perp}_{\alpha}$ at the boundary one can uniquely determine the boundary current.

It is now easy to extract the boundary action corresponding to the gauge field $A_{\mu}$. 
On-shell evaluation of the first part of \eqref{mwactiontotal} using integration by parts together with the boundary condition \eqref{bdyc} gives us
\begin{equation}
-\frac{1}{2}\int d^d x \sqrt{-g}\,F^{\mu\nu}\partial_\mu A^{\perp}_\nu= -\frac{1}{2}\int d^{d}x\,\partial_\mu\left(\sqrt{-g}F^{\mu\nu}A^{\perp}_\nu\right)=-\frac{1}{2}\int_{\partial\mathcal{R}}d^{d-1}x\,j^\alpha A^{\perp}[j]_\alpha,
\end{equation}
The second part of \eqref{mwactiontotal} is given by
\begin{equation}
\int_{\partial\mathcal{R}}d^{d-1}x\,j^\alpha \left(A^{\perp}[j]_\alpha+\partial_\alpha \phi\right).\label{japhi}
\end{equation}
We now combine these two parts and obtain the full boundary action
\begin{equation}
S^\text{bdy}\left[j,\phi\right]=\int d^{d-1}x\, \left(\frac{1}{2}j^\alpha A^{\perp}[j]_\alpha + j^\alpha \partial_\alpha\phi\right).\label{mwsbdyjphi}
\end{equation}
The boundary partition function will be
\begin{align}
    \mathcal{Z}_{\rm{bdy}}&=\int [Dj][D\phi]e^{i S_{\rm{bdy}}}.
\end{align}
\subsection{Planar boundary}
We explicitly evaluate the edge mode partition function for a planar boundary which is placed at $z=0$. The boundary conditions \eqref{bdyc} read
\begin{align}
    F_{z\alpha}|_{\rm{bdy}}=j_{\alpha},\quad A^{\perp}_z=0
\end{align}
The boundary condition implies that $\partial_z A^{\perp}_{\alpha}=j_{\alpha}$. Therefore, it is explicit that there is a linear isomorphism between $A^{\perp}_{\alpha}$ and $j_{\alpha}$. We now expand the transverse gauge field $A^{\perp}_{\alpha}$ in Fourier modes along the $z$ direction and the boundary action can be expressed as
\begin{align}
    S^\text{bdy}=\int d^{d-1}x\,i k_z \left(\frac{1}{2}A^{\perp\alpha} A^{\perp}_\alpha +A^{\perp\alpha} \partial_\alpha\phi\right)|_{z=0}.
\end{align}
Therefore the edge mode partition function can be evaluated as
\begin{align}
   \mathcal{Z}_{\rm{bdy}}&=|J|\int [DA^{\perp}_{\mu}][D\phi]e^{i S^\text{bdy} }\nonumber\\
   &=|J|\int [D\phi]e^{\frac{i}{2}\int d^{d-1}x\hspace{0.4cm}(\partial^{\alpha}\phi\partial_{\alpha}\phi})\nonumber\\
   &=\frac{|J|}{\sqrt{i k_z}}\frac{1}{\sqrt{\rm{det}(-i\Delta_0)}}.
\end{align}
$\Delta_0$ represents the Laplacian of massless scalar field evaluated at the boundary.
Here $|J|$ represents the Jacobian which comes from the linear transformation of the boundary current and the transverse gauge field at the boundary. The path integral over $A^{\perp}_{\alpha}$ gives us a constant overall factor which we absorb in the Jacobian factor. 

Finally, the edge partition function of free Maxwell field is described by the free energy of a scalar along the boundary. Note that, at a constant time slide the boundary manifold would be co-dimension-2 space and the boundary path integral would result in a massless scalar free energy along the boundary which is in agreement with the operator formalism.
\subsection{Spherical boundary} Let us now derive the edge mode action for a free Maxwell field across a spherical boundary. The boundary action is given by
\begin{align}
    S^\text{bdy}\left[j,\phi\right]=\int\sqrt{g} d^{d-1}x\, \left(\frac{1}{2}j^\alpha A[j]^{\perp}_\alpha + j^\alpha \nabla_\alpha\phi\right)
\end{align}
We vary the action with respect to the boundary current $j^{\mu}$ and obtain a condition that we impose to obtain the on-shell action.
\begin{align}
    \frac{1}{2}A^{\perp}_{\mu}+\nabla_{\mu}\phi=0
\end{align}
Note that here $x_{\mu}$ represents only the boundary coordinates and the boundary is placed at $r=R$.
The boundary condition \eqref{bdyc} reads
\begin{align}
    \left.\left(\sqrt{-g}  F^{r\alpha}\right)\right|_\text{bdy}= j^\alpha, \quad\quad A^{\perp r}|_{\rm{bdy}}=0
\end{align}
We now substitute the boundary current in terms of the boundary gauge potential and obtain
\begin{align}
     S^\text{bdy}\left[A,\phi\right]&=\frac{1}{2}\int d^{d-1}x\left(\partial_r A^{\perp\mu}A^{\perp}_{\mu}-\partial_r\phi\nabla^{\mu}\nabla_{\mu}\phi\right)|_{r=R}
\end{align}

Using the boundary condition, we obtain the boundary action
\begin{align}
    S^\text{bdy}\left[j,\phi\right]&=\frac{1}{2}\int d^{d-1}x\partial_r A^{\perp\mu}A^{\perp}_{\mu}|_{r=R}-\int d^{d-1}x \partial_r \phi\nabla^{\mu}\nabla_{\mu}\phi |_{r=R}
\end{align}
It is clear that the path integral over the boundary gauge field $A^{\perp}_{\mu}$ gives an overall constant. We now path integrate over the longitudinal modes $\phi$ along the boundary and obtain the edge partition function.
\begin{align}\label{0a}
    \mathcal{Z}_{\rm{edge}}&=\left(-\det(i \Delta_0)\right)^{-\frac{1}{2}}
\end{align}
Note that, the derivative operator in the radial coordinate is evaluated at $r=R$. Therefore, evaluation of the boundary path integral will involve the scalar Laplacian only in the boundary coordinates. This agrees with the computation of the edge modes of the free Maxwell field in the operator formalism \eqref{fm} at a constant time slice.
\section{Edge mode action of $p$-forms}\label{sec4}
We will derive the edge mode action for the $ 2$ form explicitly. We consider a free 2-form action on a Lorentzian manifold $\mathcal{R}\cup \bar{\mathcal{R}}$ without boundaries.
\begin{align}\label{act2}
    S&=-\frac{1}{12}\int _{\mathcal{R}\cup \bar{\mathcal{R}}} d^dx F_{\mu\nu\rho}F^{\mu\nu\rho}.
\end{align}
 The field strength $F_{\mu\nu\rho}$ is defined as
 \begin{align}  F_{\mu\nu\rho}&=\partial_{\mu}A_{\nu\rho}+\partial_{\nu}A_{\rho\mu}+\partial_{\rho}A_{\mu\nu},
 \end{align}and the action is invariant under the gauge transformation
 \begin{align}
     \delta A_{\mu\nu}&=\partial_{\mu}\epsilon^{\perp}_{\nu}-\partial_{\nu}\epsilon^{\perp}_{\nu}, \quad\quad\partial_{\mu}\epsilon^{\perp\mu}=0.
 \end{align}
 Note that, the longitudinal part of the gauge parameter does not appear in the gauge transformation.
 Gauge fields $A_{\mu\nu}$ and $\bar{A}_{\mu\nu}$ are defined on $\mathcal{R}$ and $\bar{\mathcal{R}}$ respectively. The full path integral is split into two parts and glued together at the boundary $\partial\mathcal{R}.$
\begin{equation}
\int \left[\mathcal{D} A_{\mu\nu}\right]\left[\mathcal{D} \bar{A}_{\mu\nu} \right] \prod_{x \in \partial \mathcal{R}} \delta\left(A_{\mu\nu}-\bar{A}_{\mu\nu}\right) e^{i S_A} e^{i S_{\bar{A}}}.
\end{equation}
The delta-function constraint can be expressed as a path integral over antisymmetric boundary current $j^{\mu\nu}$.
\begin{equation}
\prod_{x \in \partial \mathcal{R}} \delta\left(A_{\mu\nu}-\bar{A}_{\mu\nu}\right) = \int \left[\mathcal{D}j^{\mu\nu}\right] e^{\frac{i}{2}\int_{\partial \mathcal{R}}d^{d-1}x\, j^{\mu\nu} ( A_{\mu\nu}-\bar{A}_{\mu\nu})}\label{2fdeltapij}
\end{equation}
Therefore, total action is given by
\begin{align}
\nonumber S= S_A + S_{\bar{A}} =&-\frac{1}{12}\int _{\mathcal{R}}d^d x\sqrt{-g}\,F^{\mu\nu\rho}F_{\mu\nu\rho}\\&+\frac{1}{2}\int_{\partial \mathcal{R}}d^{d-1}x\, j ^{\mu\nu} ( A_{\mu\nu}-\bar{A}_{\mu\nu})-\frac{1}{12}\int _{\bar{\mathcal{R}}}d^d x\sqrt{-g}\,\bar{F}^{\mu\nu\rho}\bar{F}_{\mu\nu\rho},
\end{align}
Let us now focus on the action on the manifold $\mathcal{R}.$
\begin{align}\label{act21}
    S_A&=-\frac{1}{12}\int _{\mathcal{R}}d^d x\sqrt{-g}\,F^{\mu\nu\rho}F_{\mu\nu\rho}+\frac{1}{2}\int_{\partial \mathcal{R}}d^{d-1}x\, j ^{\mu\nu} A_{\mu\nu}
\end{align}
It is clear that under gauge transformation, the first part of the action \eqref{act21} does not change. But the second part of the action is not invariant under gauge transformation. Therefore under a gauge transformation, the action becomes
\begin{align}
     S_A&=-\frac{1}{12}\int _{\mathcal{R}}d^d x\sqrt{-g}\,F^{\mu\nu\rho}F_{\mu\nu\rho}+\frac{1}{2}\int_{\partial \mathcal{R}}d^{d-1}x\, j ^{\mu\nu}\left( A^{\perp}_{\mu\nu}+\partial_{\mu}\epsilon^{\perp}_{\nu}-\partial_{\nu}\epsilon^{\perp}_{\nu}\right)
\end{align}
In the presence of a boundary, gauge invariance is broken again and the transverse part of the gauge parameter $\epsilon^{\perp}_{\mu}$ appears as physical degrees of freedom along the boundary. To obtain the partition function, one requires a path integral over transverse $A^{\perp}_{\mu\nu}$, gauge parameter $\epsilon^{\perp}_{\mu}$ and the boundary current $j_{\mu\nu}$. The path integral over transverse gauge field $A^{\perp}_{\mu\nu}$ will give us the bulk partition function. 
\begin{align}
    \mathcal{Z}[j,\epsilon]&=\int_{\mathcal{R}}[DA^{\perp}_{\mu\nu}]e^{i S_A}=\mathcal{Z}_{\rm{bulk}}\times \mathcal{Z}_{\rm{bdy}}.
\end{align}
After stripping off the bulk partition function, we path integrate over the boundary action and the boundary gauge field $A^{\perp}_{\mu\nu}$ is evaluated on-shell. At this stage, we impose the boundary condition across the boundary.
\begin{align}\label{bdy2}
    \left.\left(\sqrt{-g} n_\mu F^{\mu\nu\alpha}\right)\right|_\text{bdy}= j^{\nu\alpha}, \quad\quad n_{\alpha}A^{\perp\mu\alpha}|_{\rm{bdy}}=0
\end{align}
Due to the boundary condition \eqref{bdy2}, we see that there is an isomorphism between the transverse component of the gauge field and the current at the boundary.

We now separate out the boundary action. Let us first focus on the first part of \eqref{act21}
\begin{align}
&-\frac{1}{12}\int d^d x \sqrt{-g}\,F^{\mu\nu\rho}\left(\partial_\mu A^{\perp}_{\nu\rho}+\partial_{\nu}A^{\perp}_{\rho\mu}+\partial_{\rho}A^{\perp}_{\mu\nu}\right)= -\frac{1}{12}\int d^{d}x\,\partial_\mu\left(\sqrt{-g}F^{\mu\nu\rho}A^{\perp}_{\nu\rho}\right)\nonumber\\
&\quad\quad\quad\quad\quad\quad\quad\quad\quad-\frac{1}{12}\int d^{d}x\,\partial_\nu\left(\sqrt{-g}F^{\mu\nu\rho}A^{\perp}_{\rho\mu}\right)-\frac{1}{12}\int d^{d}x\,\partial_\rho\left(\sqrt{-g}F^{\mu\nu\rho}A^{\perp}_{\mu\nu}\right)\nonumber\\
&\quad\quad\quad\quad\quad\quad\quad\quad\quad\quad\quad\quad\quad\quad\quad\quad\quad\quad\quad\quad\quad=-\frac{1}{4}\int d^{d-1}x\quad j^{\mu\nu}A^{\perp}_{\mu\nu}
\end{align}
Combining with the second part of the action, we obtain
\begin{align}
    S^{\rm{bdy}}[j,\epsilon]&=-\frac{1}{4}\int d^{d-1}x \quad j^{\mu\nu}\left(A^{\perp}_{\mu\nu}+2(\partial_{\mu}\epsilon^{\perp}_{\nu}-\partial_{\nu}\epsilon^{\perp}_{\mu})\right).
\end{align}
Therefore the edge partition function for the 2-form field will be
\begin{align}
    \mathcal{Z}_{\rm{edge}}&=\int [Dj][D\epsilon] e^{i S_{\rm{bdy}}}
\end{align} Let us now generalize our analysis for arbitrary antisymmetric $p$-form gauge potential. We follow a similar procedure to obtain the action in the manifold $\mathcal{R}.$
\begin{align}
     S_A&=-\frac{1}{2(p+1)!}\int _{\mathcal{R}}d^d x\sqrt{-g}\,F^{\mu_1\cdots \mu_p}F_{\mu_1\cdots\mu_p}\nonumber\\&+\frac{1}{2}\int_{\partial \mathcal{R}}d^{d-1}x\, j ^{\mu_1\cdots\mu_p}\Big( A^{\perp}_{\mu_1\cdots\mu_p}+\partial_{\mu_1}\epsilon^{\perp}_{\mu_1\cdots\mu_{p-1}}+\cdots+(-1)^{p-1}\partial_{\mu_{p-1}}\epsilon^{\perp}_{\mu_{1}\cdots\mu_{p-2}}\Big)
\end{align}
To evaluate the boundary action, we focus on the surface term of the action and impose the following boundary conditions 
\begin{align}\label{bdycp}
      \left.\left(\sqrt{-g} n_{\mu_1} F^{\mu_1\cdots\mu_{p+1}}\right)\right|_\text{bdy}= j^{\mu_1\cdots\mu_p}, \quad\quad n_{\mu_1}A^{\perp \mu_1\cdots\mu_p}|_{\rm{bdy}}=0
\end{align}
Following the similar procedure developed for the $2$-form, we obtain the action of the edge modes of free $p$-forms
\begin{align}
    S^{\rm{bdy}}[j,\epsilon]&=-\frac{1}{2(p+1)!}\int d^{d-1}x  j^{\mu_1\cdots\mu_{p}}\Big(A^{\perp}_{\mu_1\cdots\mu_{p}}\nonumber\\
&\quad\quad\quad\quad\quad\quad\quad\quad\quad\quad+2(\partial_{\mu_1}\epsilon^{\perp}_{\mu_2\cdots\mu_{p-1}}+\cdots+(-1)^{p-1}\partial_{\mu_{p-1}}\epsilon^{\perp}_{\mu_1\cdots\mu_{p-2}})\Big). 
\end{align}
So the boundary action of free $p$-forms involves the transverse gauge field as well as the gauge parameter $\epsilon$. In the next section, we will evaluate the boundary path integral to obtain the partition function of the edge modes of free $p$-forms.
\subsection{Planar boundary}
We now evaluate the edge mode partition function of $p$-forms across a planar boundary which is placed at $z=0$. The boundary conditions \eqref{bdycp} read
\begin{align}\label{dycp}
    F_{z\mu_1\cdots \mu_p}|_{\rm{bdy}}=j_{\mu_1\cdots \mu_p},\quad A^{\perp}_{z\mu_1\cdots\mu_{p-1}}=0.
\end{align}
The boundary condition implies that $\partial_z A^{\perp}_{\mu_1\cdots\mu_{p}}=j_{\mu_1\cdots\mu_p}$. Therefore, it is explicit that there is a linear isomorphism between gauge potential $A^{\perp}_{\mu_1\cdots\mu_p}$ and the boundary current $j_{\mu_1\cdots\mu_p}$. It is now good to write the boundary action in terms of the Fourier modes along the $z$ direction
\begin{align}
    S^\text{bdy}=\int d^{d-1}x\,i k_z \left(\frac{1}{2}A^{\perp\mu_1\cdots\mu_p} A^{\perp}_{\mu_1\cdots\mu_p}+A^{\perp\mu_1\cdots\mu_p} (\partial_{\mu_1}\epsilon^{\perp}_{\mu_2\cdots\mu_{p-1}}+\rm{cyclic}\right)|_{z=0}.
\end{align}
Since there is a linear isomorphism between the gauge potential and the boundary current, we can finally obtain the boundary action in terms of the gauge potential which obeys the boundary condition \eqref{bdycp}. Therefore, the path integral will be done over the boundary gauge potential and the gauge parameter $\epsilon$. There will also be a constant Jacobian due to the transformation from the boundary current to the gauge potential. It is clear that the path integral over the gauge parameters will give us transverse $(p-1)$-form Laplacian on the boundary. To be explicit, let us evaluate the boundary path integral for $2$-forms.
\begin{align}
     \mathcal{Z}_{\rm{bdy}}&=|J|\int [DA^{\perp}_{\mu\nu}][D\epsilon^{\perp}_{\mu}]e^{i S^\text{bdy} },\nonumber\\
\end{align}
where the boundary action in this case will be
\begin{align}
   S^\text{bdy}=\int d^{d-1}x\,i k_z \left(\frac{1}{2}A^{\perp\mu\nu} A^{\perp}_{\mu\nu}+A^{\perp\mu\nu} (\partial_{\mu}\epsilon^{\perp}_{\nu}-\partial_{\nu}\epsilon^{\perp}_{\mu}\right)|_{z=0}  
\end{align}

The path integral over the gauge potential will give us
\begin{align}
    \mathcal{Z}_{\rm{edge}}&=\frac{|J|}{\sqrt{ik_z}}\int [D\epsilon^{\perp}_{\mu}]e^{-\int d^{d-1}x\quad\partial^{\nu}\epsilon^{\perp\mu}\partial_{\nu}\epsilon^{\perp}_{\mu}}\nonumber\\
    &=\frac{|J|}{\sqrt{ik_z}}\frac{1}{\sqrt{\det \Delta^{(-i\perp}_1)}}
\end{align}
Therefore, for free $2$-forms, the boundary partition function or the edge free energy becomes free energy of the transverse $1$-form on the boundary. We can now easily generalize it to antisymmetric $p$-forms. Note that, the gauge parameter in this case is transverse $(p-1)$-forms, and the path integral over the gauge parameter will lead us to the determinant of the Laplacian of co-exact $(p-1)$-forms.
\begin{align}
      \mathcal{Z}_{\rm{edge}}&=\frac{|J|}{\sqrt{ik_z}}\int [D\epsilon^{\perp}_{\mu_1\cdots\mu_{p-1}}]e^{-\int d^{d-1}x\quad\partial^{\nu}\epsilon^{\perp\mu_1\cdots \mu_{p-1}}\partial_{\nu}\epsilon^{\perp}_{\mu_1\cdots\mu_{p-1}}}\nonumber\\
    &=\frac{|J|}{\sqrt{ik_z}}\frac{1}{\sqrt{\det \Delta^{\perp}_{p-1}}},
\end{align}
where $\Delta^{\perp}_{p-1}$ is the transverse Laplacian of $(p-1)$-forms with the boundary co-ordinates.  
\subsection{Spherical boundary}Let us now derive the partition function of edge modes of $p$-forms across a spherical boundary. The boundary is placed at $r=R$. The boundary action is given by
\begin{align}
    S^\text{bdy}\left[j,\epsilon\right]=\int\sqrt{g} d^{d-1}x\, \left(\frac{1}{2}j^{\mu_1\cdots\mu_p} A^{\perp}[j]_{\mu_1\cdots\mu_p} + j^{\mu_1\cdots\mu_{p}} (\nabla_{\mu_1}\epsilon^{\perp}_{\mu_2\cdots\mu_{p-1}}+\rm{cyclic})\right).
\end{align}
We vary the action with respect to the boundary current $j^{\mu_1\cdots\mu_{p}}$ and obtain a condition that we impose to obtain the on-shell action.
\begin{align}
    \frac{1}{2}A_{\mu_1\cdots\mu_p}+(\nabla_{\mu_1}\epsilon_{\mu_2\cdots\mu_{p-1}}+\rm{cyclic})=0.
\end{align}
Note that here $x_{\mu}$ represents only the boundary coordinates.
The boundary condition \eqref{bdycp} reads
\begin{align}
    \left.\left(\sqrt{-g}  F^{r\mu_1\cdots\mu_p}\right)\right|_\text{bdy}= j^{\mu_1\cdots\mu_p}, \quad\quad A^{\perp }_{r\mu_1\cdots\mu_{p-1}}|_{\rm{bdy}}=0.
\end{align}
The boundary conditions \eqref{bdycp} imply that there is an isomorphism between the transverse gauge potential and the boundary current. We first evaluate the boundary path integral for 2-form which will help us to generalize it to $p$-forms.
The boundary action for 2-forms on the sphere is given by
\begin{align}
     S^\text{bdy}=\int \sqrt{g} d^{d-1}x\ \left(\frac{1}{2}A^{\perp\mu\nu} A^{\perp}_{\mu\nu}+A^{\perp\mu\nu} (\nabla_{\mu}\epsilon^{\perp}_{\nu}-\nabla_{\nu}\epsilon^{\perp}_{\mu})\right)|_{r=R}.
\end{align}
To obtain the boundary partition function, we first path integrate over the transverse gauge field and obtain
\begin{align}
    \mathcal{Z}_{\rm{edge}}&=\int [D\epsilon^{\perp}_{\mu}]\exp\Big[i\int\sqrt{g} d^{d-1}x\left(\nabla_{\mu}\epsilon^{\perp\nu}\nabla^{\mu}\epsilon^{\perp}_{\nu}+\nabla_{\mu}\epsilon^{\perp\nu}\nabla^{\nu}\epsilon^{\perp}_{\mu}\right)].
\end{align}
The path integral can now be performed easily by integrating by parts of the boundary action and the first term will be spin-1 Laplacian and the second term will give us the curvature-induced mass term on the sphere. 
Therefore it is clear that the path integral over the gauge parameter $\epsilon^{\perp}_{\mu}$ will result in the Hodge-de Rham Laplacian of $1$-form on the boundary. 
\begin{align}
\mathcal{Z}_{\rm{edge}}&=\frac{1}{\sqrt{\det(-i\Delta_{(1)\rm{HdR}})}}.
\end{align}
It is not very hard to generalize it to $p$-form. The structure we obtain in 2-form explicitly will help us to compute the boundary path integral in $p$-forms.
The boundary action is given by
\begin{align}
     S^\text{bdy}=\int \sqrt{g} d^{d-1}x\ \left(\frac{1}{2}A^{\perp\mu_1\cdots\mu_p} A^{\perp}_{\mu_1\cdots\mu_p}+A^{\perp\mu_1\cdots\mu_p} (\nabla_{\mu_1}\epsilon^{\perp}_{\mu_2\cdots\mu_{p-1}}+\rm{cyclic}\right)|_{r=R}.
\end{align}
To obtain the boundary partition function, we first path integrate the transverse gauge field. The path integral over the transverse gauge field will give us a quadradic boundary action in terms of the gauge parameter $\epsilon^{\perp}$ which is the square of the second term in the parenthesis. Note that, the gauge parameter is a co-exact $(p-1)$-form, and the path integral over it will give us the Hodge-de Rham Laplacian of co-exact $(p-1)$-form on the boundary.
\begin{align}\label{pap}
\mathcal{Z}_{\rm{edge}}&=\frac{1}{\sqrt{\det(-i\Delta_{(p-1)\rm{HdR}})}}.
\end{align}
It is important to note that, this boundary path integral procedure agrees with the direct evaluation of the edge modes of $p$-forms in section \eqref{sec2p}.
\section{Relation with the edge partition function on sphere}\label{sec5}
Free energies of free quantum fields on spheres, anti-de Sitter spaces, and de Sitter spaces have been quite useful in the evaluation of entanglement entropies, quantum corrections to black hole entropy, and holography. There is a recent development where free energies are expressed as integrals
over Harish-Chandra characters \cite{Anninos:2020hfj,David:2021wrw}. In this section, we will relate the edge partition function of $p$-forms on the sphere with the entanglement entropy of edge modes. 

Let us briefly review the structure of the bulk-edge decomposition of the free energies of $p$-forms on the sphere \cite{David:2020mls}.  The expression  for the one loop path integral of free $p$-forms on
 $S^{d}$ can be written as follows 
 \begin{eqnarray}
 \log {\cal Z}  &=& \log {\cal Z}_G +  \log {\cal Z}_{{\rm chr} }, \\ \nonumber
\log {\cal Z}_{{\rm chr} } &=&   \int_0^\infty  \frac{dt}{2t}\frac{ 1+ e^{-t} }{ 1-  e^{-t} }
 \left( \chi_{{\rm bulk }} ( t) - \chi_{{\rm edge}} ( t) 
 \right).
 \end{eqnarray}
 Here ${\cal Z}_G $ contains  the information about the dimensionless 
 coupling constant of the theory.  It also contains the volumes of the gauge groups of $p$-froms in the one-loop partition function. 
 The second part of the free energy has an interesting structure.  $ \log {\cal Z}_{{\rm chr} }$  exhibits the bulk-edge decomposition when expressed in terms of the characters.
  The bulk contribution  can be written as 
 an integral transform of  Harish-Chandra characters of the group $SO(1, d) $, while the 
 edge term can be written as an integral transform of the Harish-Chandra character in two lower dimensions which is $d\rightarrow d-2$. 
 Let us now understand this explicitly. The partition function or the path integral of gauge fixed free $p$-form field on $S^{d}$ is given by
\cite{Obukhov:1982dt,Copeland:1984qk,Cappelli:2000fe}
\begin{align} \label{gfpform}
    \mathcal{Z}_p[S^{d}]
    &=\Big[\frac{1}{\det_T \Delta_p}\frac{\det_T\Delta_{p-1}}{\det_T\Delta_{p-2}}\cdots\big(\frac{\det_T\Delta_1}{\det'\Delta_0}{\rm Vol }S^{d}\big)^{(-1)^p}\Big]^{\frac{1}{2}}.
\end{align}
$\det_T\Delta_p$ denotes the determinant of Hodge-de Rham Laplacian of the co-exact or transverse $p$-forms.
The prime in 
$\det'\Delta_0$  refers to the determinant of massless scalar with the zero modes removed. It is clear that the key ingredient in this partition function is the determinant of Hodge-de Rham Laplacian of the co-exact or transverse $p$-forms. Let us evaluate it carefully.
\begin{eqnarray}\label{detegi}
-\frac{1}{2} \log ( {\rm det}_T \Delta_p^{S^{d}} )  = - \sum_{n=1}^\infty 
\frac{1}{2} g_{n}^{(p)} \log ( \lambda_{n }^{(p)} ) .
\end{eqnarray}
The eigenvalues $\lambda_{n}^{(p)}$ and the degeneracies $g_{n}^{(p)}$ of the Hodge-de Rham Laplacian 
of  co-exact $p$-forms on $S^{d}$ are given by \cite{Copeland:1984qk}
\begin{equation} \label{degeig}
    \begin{split}
        \lambda_{n}^{(p)}&=(n+p)(n+d-p-1),\\
             g_{n}^{(p)} &=\frac{(2n+d-1)\Gamma[n+d]}{\Gamma[p+1]\Gamma[d-p]\Gamma[n](n+p)(n+d-p-1)}.
    \end{split}
\end{equation}
At this stage, one replaces the logarithm with the following identity
\begin{equation}\label{logiden}
-\log y  = \int_0^\infty \frac{d\tau}{\tau} ( e^{-y \tau} - e^{-\tau} ) .
\end{equation}
Substituting this identity in (\ref{detegi}), we obtain 
\begin{eqnarray}\label{detegi2}
-\frac{1}{2} \log ( {\rm det}_T \Delta_p^{S^{d+1}} )  = \int_0^\infty \frac{d\tau}{2\tau} 
\left( \sum_{n=1}^\infty  g_{ n}^{(p)} ( e^{-\tau \lambda_{n}^{(p)}  } -   e^{-\tau})  \right) .
\end{eqnarray}.
We now then perform the $\tau$ integral and sum over the eigenmodes and use the Hubbard-Stratonovich trick to obtain the determinant of co-exact $p$-forms in terms of the characters. Interestingly, the determinant can be expressed as an integral transform of the characters of co-exact $p$-forms. The explicit expression is given by \cite{David:2021wrw}
\begin{align}
    \chi^{dS}_{(d,p)}(t) &=\binom{d-1}{p}\frac{e^{-t(d-p-1)}+e^{-tp}}{(1-e^{-t})^{d-1}}.
\end{align}
Now the determinant of the co-exact $p$-form becomes
 \begin{align}\label{chrintrep}
  -\frac{1}{2} \log ( {\rm det}_T \Delta_p^{S^{d} })  
  &=\int_{0}^{\infty}\frac{dt}{2t}\frac{ 1+ e^{-t}}{ 1- e^{-t} }
  \left(\sum_{i=0}^{p}(-1)^i \chi^{dS}_{(d-1-2i,p-i)}(t)\right).
\end{align}
In this expression, the term $i =0$, is the `naive' bulk character   \cite{Anninos:2020hfj}
of the co-exact $p$-form. All terms $i\geq 1$ constitute the `naive' edge characters.

Let us now write down the expressions for bulk and edge characters of co-exact 1-form on $S^d$.
\begin{align}
    \chi^{\rm{bulk}}_{(d,1)}=(d-1)\frac{e^{-t(d-2)}+e^{-t}}{(1-e^{-t})^{d-1}},\quad \chi^{\rm{edge}}_{(d,1)}=\frac{e^{-t(d-3)}+1}{(1-e^{-t})^{d-3}}.
\end{align}
Therefore, the edge determinant of the co-exact $1$-form can be written as
\begin{align}
      -\frac{1}{2} \log ( {\rm det}_T \Delta_1^{S^{d} })|_{\rm{edge}}&=-\int_{0}^{\infty}\frac{dt}{2t}\frac{ 1+ e^{-t}}{ 1- e^{-t} }\frac{e^{-t(d-3)}+1}{(1-e^{-t})^{d-3}}.
\end{align}
In \cite{David:2021wrw}, it has been identified as a partition function of massless scalar field or $0$-form on $S^{d-2}$. We can also see this explicitly from the operator formalism computation \eqref{HdR1} as well as from the boundary action in \eqref{0a}. Therefore, we identify the logarithmic divergent term of entanglement entropy edge modes in free Maxwell field with the log divergent part of the edge partition function on the sphere.

Let us now verify this in higher forms. For co-exact $p$-forms on $S^d$, the edge partition function is given by
\begin{align}\label{edgech}
     -\frac{1}{2} \log ( {\rm det}_T \Delta_p^{S^{d} })|_{\rm{edge}}&=-\int_{0}^{\infty}\frac{dt}{2t}\frac{ 1+ e^{-t}}{ 1- e^{-t} }
  \left(\sum_{i=1}^{p}(-1)^i \chi^{dS}_{(d-1-2i,p-i)}(t)\right),\nonumber\\
  &=\int_{0}^{\infty}\frac{dt}{2t}\frac{ 1+ e^{-t}}{ 1- e^{-t} }
  \left(\sum_{i'=0}^{p}(-1)^{i'} \chi^{dS}_{(d-3-2i',p-1-i')}(t)\right).
\end{align}
Comparing \eqref{edgech} with \eqref{chrintrep}, we find that the edge character of co-exact $p$-forms on $S^d$ is the same as partition (bulk and edge) function of $(p-1)$-forms on $S^{d-2}$. It is clear that the free energy of co-exact $(p-1)$-from on $S^{d-2}$ can be identified with the edge partition functions of co-exact $p$-forms on $S^d$. Therefore, we find an agreement with the operator formalism \eqref{ppart} and with the boundary path integral computation of edge modes \eqref{pap}.

In \cite{David:2020mls}, the entanglement entropy of conformal $p$-forms has been evaluated using a conformal mapping procedure \cite{Casini:2011kv}. The conformal mapping procedure enables us to compute the entanglement entropy of the ground state of conformal fields from the partition function on the hyperbolic cylinder. In \cite{David:2021wrw}, it was shown that the partition function of conformal $p$-forms on hyperbolic cylinder agrees with the bulk partition function on sphere and misses out on the edge partition function. In this paper, we explicitly show that the entanglement entropy of edge modes of $p$-forms agrees with the edge partition function. Therefore, for conformal $p$-forms the extractable entanglement entropy can be evaluated from the bulk partition function, and the edge mode entanglement entropy can be extracted from the edge partition function on the sphere. However, for free $p$-forms including non-conformal ones, the extractable entanglement entropy has not been evaluated. It is best to proceed with mode quantization and relate the Hamiltonian with the scalar Hamiltonian with certain modes removed, as it was done for free Maxwell field and graviton in $d=4$ dimension \cite{Benedetti:2019uej}.

\section{Discussion}

In this paper, we isolate the entanglement entropy of edge modes of free $p$-forms corresponding to the electric center across a spherical entangling surface. Using operator formalism as well as from the boundary path integral, we find that the edge mode of free $p$-form field across a sphere can be understood as the free energy of $(p-1)$-form on codimension-2 sphere. Interestingly, the logarithmic divergent part of the entanglement entropy of edge modes coincides with the log divergent part of the edge partition of $p$-forms when expressed in terms of the Harish-Chandra character.

The extractable part of the entanglement entropy of conformal $p$-forms across a spherical entangling surface was evaluated in \cite{David:2020mls} using a conformal mapping procedure. The conformal mapping procedure enables us to evaluate the entanglement entropy of a ground state in conformal field theory from the thermal partition function on the hyperbolic cylinder \cite{Casini:2011kv}. The logarithmic divergent part of the extractable entanglement entropy of conformal $p$-forms coincides with the bulk partition function of the same when expressed in terms of its Harish-Chandra character \cite{David:2021wrw}. However this conformal mapping prescription works only for the conformal fields and therefore, one requires to develop a mode quantization procedure to isolate the extractable part for free $p$-forms in arbitrary dimension. We wish to investigate this in the near future. 

It will be nice to have the path integral prescription of the edge modes in gravity. Recently, there has been recent progress in understanding the gravitational edge modes in terms of the algebra of the boundary charges \cite{Freidel:2020xyx,Freidel:2020svx,Freidel:2020ayo}.\footnote{There are also recent developments in understanding gravitational edge modes in lower dimensions \cite{Joung:2023doq, Mertens:2022ujr}.} In \cite{David:2022jfd}, the entanglement entropy of graviton has been evaluated in $d=4$ dimension across a spherical entangling surface. We wish to identify the entanglement entropy of these edge modes in terms of the algebra of the boundary charges.

We understand that the presence of a boundary breaks the gauge symmetry in a gauge theory and as a result, the longitudinal modes appear to be physical on the boundary. This is reflected in the structure of the edge mode partition function. However, it is not clear why the edge partition function on the sphere agrees with the entropy of the edge modes corresponding to the electric centre. We wish to have a first-principle derivation that clarifies this precise agreement.

The bulk-edge decomposition in the partition function on sphere has also been observed in conformal higher derivative gauge fields and conformal higher spin theories \cite{Mukherjee:2021rri,Mukherjee:2021alj}. It will be nice to evaluate the edge mode entanglement in these theories using operator formalism as well as from boundary path integral and check whether the agreement holds true even in these cases. These exercises will reveal the nature of the edge modes in higher spin theories. It will also be nice to explore to investigate the edge mode structure in excited states of gauge theories \cite{David:2022czg,Mukherjee:2022jac}.
\section{Acknowledgement}The author wishes to thank Justin David and Onkar Parrikar for the discussions.
\bibliographystyle{JHEP}
\bibliography{reference.bib}
\end{document}